\documentclass[12pt]{article}
\usepackage{pgf,authblk}
\usepackage{amsmath,amssymb}
\usepackage{pgfpages,graphicx}
\usepackage{hyperref} 
\usepackage{epsfig,multicol,mcite,amsmath}
\usepackage{marvosym}
\begin{document}                                     
\title{DISPred - a program to calculate deep inelastic scattering cross sections\\ v1.0 }
\author{J. Ferrando}
\affil{University of Oxford}

\maketitle

\abstract{
A new package, DISPred, is described. The package can be used to calculate $e^{\pm}$p deep inelastic scattering cross sections at Born level in Electroweak theory and at both leading and next-to-leading order in QCD.
}

\section{Introduction}

The package DISPred arose as a result of the need to produce 
 predictions of deep inelastic scattering (DIS) electron-proton cross
 sections at next-to-leading order (NLO) in QCD for comparison to data in
 ZEUS publications. In the current version (1.0) predictions for the following are available at both leading order (LO) and NLO in QCD:

\begin{itemize}
\item the reduced/ double-diff cross sections for neutral current (NC) and charged current (CC) DIS
\item the total cross section $\sigma_{tot}$ for NC and CC DIS
\item differential cross sections $\frac{d\sigma}{dQ^2}$,  $\frac{d\sigma}{dx}$, $\frac{d\sigma}{dy}$ for NC and CC DIS
\end{itemize}
for collisions of unpolarized beams of electrons ($e^{\pm}$) and protons.

DISPred has been tested with LHAPDF 5.8.2~\cite{Whalley:2005nh} and can produce predictions for the ZEUS-JETS~\cite{Chekanov:2005nn} or HERA0.1 parton distribution functions (PDFs) in LHAPDF in the {\tt .LHpdf} format, or any other PDF within LHAPDF in the {\tt .LHgrid} format.

It produces output in ascii text format and can also produce histograms and graphs in ROOT-based \cite{Brun:1997pa} formats.

\section{Leading Order Calculation}
The LO QCD, Born-level electroweak, cross section is calculated
according to the formulation given by Devenish and Cooper-Sarkar\cite{Devenish:2004pb}.

\subsection{Reduced and Double Differential Cross Sections}
\subsubsection{NC DIS}
The double differential cross section in NC scattering is:
\begin{equation}
\frac{\mathrm{d}^2\sigma^{e^{\pm}p}_{\mathrm{NC}}}{\mathrm{d}x \mathrm{d}Q^2} = \frac{2 \pi \alpha^2  Y_{+}}{xQ^4}[F_2^{\mathrm{NC}}(x,Q^2)- \frac{y^2}{Y_{+}}F_{L}^{\mathrm{NC}}(x,Q^2) \mp \frac{Y_{-}}{Y_{+}}xF_3^{\mathrm{NC}}(x,Q^2)] \label{eq:ncdisdd}.
\end{equation}

Where, as is conventional, $Q^2$ is the virtuality of the exchanged boson, $x$ is the momentum fraction of the struck parton in the infinite proton-momentum, $F^{\mathrm{NC}}_i$ are structure functions defined later, $Y_{\pm}=  1\pm(1-y)^2$ and $y$ is the inelasticity of the electron.

For the leading order calculation the structure functions are defined as follows:

\begin{equation}
F^{\mathrm{NC}}_2= \sum\limits_i A_i^0(Q^2)(xq_i(x,Q^2) + x\bar{q}_i(x,Q^2)) ;
\end{equation}

\begin{equation}
F^{\mathrm{NC}}_L=0 ;
\end{equation}

\begin{equation}
xF^{\mathrm{NC}}_3= \sum\limits_i B_i^0(Q^2)(xq_i(x,Q^2) - x\bar{q}_i(x,Q^2)) ; ;
\end{equation}

where $A_i$ and $B_i$ can be expressed in terms of the NC vector and  axial-vector electroweak couplings to the quarks (electron) $v_i$ ($v_e$) and $a_i$ ($a_e$) and quark charge $e_i$ as

\begin{equation}
A_i^0= e^2_i - 2e_i v_i v_e P_Z(Q^2) + (v_e^2+a_e^2)(v_i^2+a_i^2)P^2_Z(Q^2)
\end{equation}

and

\begin{equation}
B_i^0= -2e_ia_ia_eP_Z(Q^2)+4a_iv_iv_ea_eP_z^2(Q^2)
\end{equation}

and

\begin{equation}
P_Z(Q^2)= \frac{Q^2}{Q^2+M_Z^2}\left(\frac{1}{\sin^2{2\theta_W}} \right)
\end{equation}

The reduced cross section for NC scattering is:
\begin{equation}
\tilde{\sigma}^{e^{\pm}p}_{\mathrm{NC}} = [F_2^{\mathrm{NC}}(x,Q^2)- \frac{y^2}{Y_{+}}F_{L}^{\mathrm{NC}}(x,Q^2) \mp \frac{Y_{-}}{Y_{+}}xF_3^{\mathrm{NC}}(x,Q^2)] \label{eq:ncdisred}.
\end{equation}

\subsubsection{CC DIS}
The double differential cross section in CC scattering is:
\begin{equation}
\frac{\mathrm{d}^2\sigma^{e^{\pm}p}_{\mathrm{CC}}}{\mathrm{d}x \mathrm{d}Q^2} = \frac{G^2_FM^2_W }{4\pi x (Q^2+M_W^2)^2}[Y_{+}F_2^{\mathrm{CC}}(x,Q^2)- {y^2}F_{L}^{\mathrm{CC}}(x,Q^2)\mp Y_{-}xF_3^{\mathrm{CC}}(x,Q^2)].
\end{equation}
Where $M_W$ is the mass of the $W$ boson, $G_F$ the Fermi coupling constant and the $F_i^{\mathrm{CC}}$ are defined at LO in QCD as\footnote{In the expressions shown above the small, Cabbibo-suppressed, contribution from the $b$-quark is neglected, it is however included in the calculation made by DISPred. In v1.0 DISPred is only suitable for use for the HERA energy regime and so top quark contributions are not included}:
\begin{eqnarray}
F_{2,e^+}^{CC}& = & x(d+s+\bar{u}+\bar{c}) \\
xF_{3,e^+}^{CC}& =& x(d+s-\bar{u}-\bar{c}) \\
F_{2,e^-}^{CC}& =& x(u+c+\bar{d}+\bar{s}) \\
xF_{3,e^-}^{CC} &= &x(u+c-\bar{d}-\bar{s}) 
\end{eqnarray}

which leads to the following expressions for the reduced cross sections:

\begin{eqnarray}
\tilde{\sigma}_{e^+}^{CC} & = & x[(1-y)^2(d+s)+\bar{u}+\bar{c}]\\
\tilde{\sigma}_{e^-}^{CC} & = & x[(1-y)^2(\bar{d}+\bar{s})+u+c]
\end{eqnarray}

\subsection{Single Differential and Total Cross Sections}

In order to calculate the single differential and total cross sections, the 
expression for the double differential cross sections is integrated over the
allowed regions of $Q^2$, $x$ and $y$ using the VEGAS~\cite{Lepage:1977sw} 
algorithm as implemented in the GNU Scientific Library\cite{gsl:3rd}. The 
number of calls used in VEGAS may be specified via the control cards. 
Differential cross sections may also be calculated at a ``point''; in this case
the width of the bin which contains the point is multiplied by a predetermined factor (which can be chosen in the control cards) to provide an approximate calculation. If no input points are specifed via cards, the option ``AUTO'' may be chosen which makes the program {\tt DISPrediction} calculate the differential cross sections at the centroid of the bin.
\section{Next-to-Leading Order Calculation}

\subsection{Reduced and Double Differential Cross Sections}
The implementation of {\sc Qcdnum} 16.13~\cite{Botje:1999dj} included in LHAPDF is used in DISPred to evaluate structure functions $F_2$, $F_L$ and $F_3$ at NLO in QCD. The prescription used by the ZEUS collaboration for the ZEUS JETS
fit~\cite{Chekanov:2002pv,Chekanov:2005nn} has been adopted. As such {\tt DISPred} can perform the QCD evolution for the ZEUS-JETS and ZEUS-S fits and use the {\tt .LHpdf} format files from LHAPDF for this. In the case of other PDFs 
{\tt DISPred} can fill a $Q^2,x$ grid for {\sc Qcdnum} using the values from the {tt .LHGrid} file. The structure functions are then generated from this grid.All other aspects of the reduced cross section cross section are the same as for the leading order case.

\begin{figure}
\includegraphics[width=9.0cm]{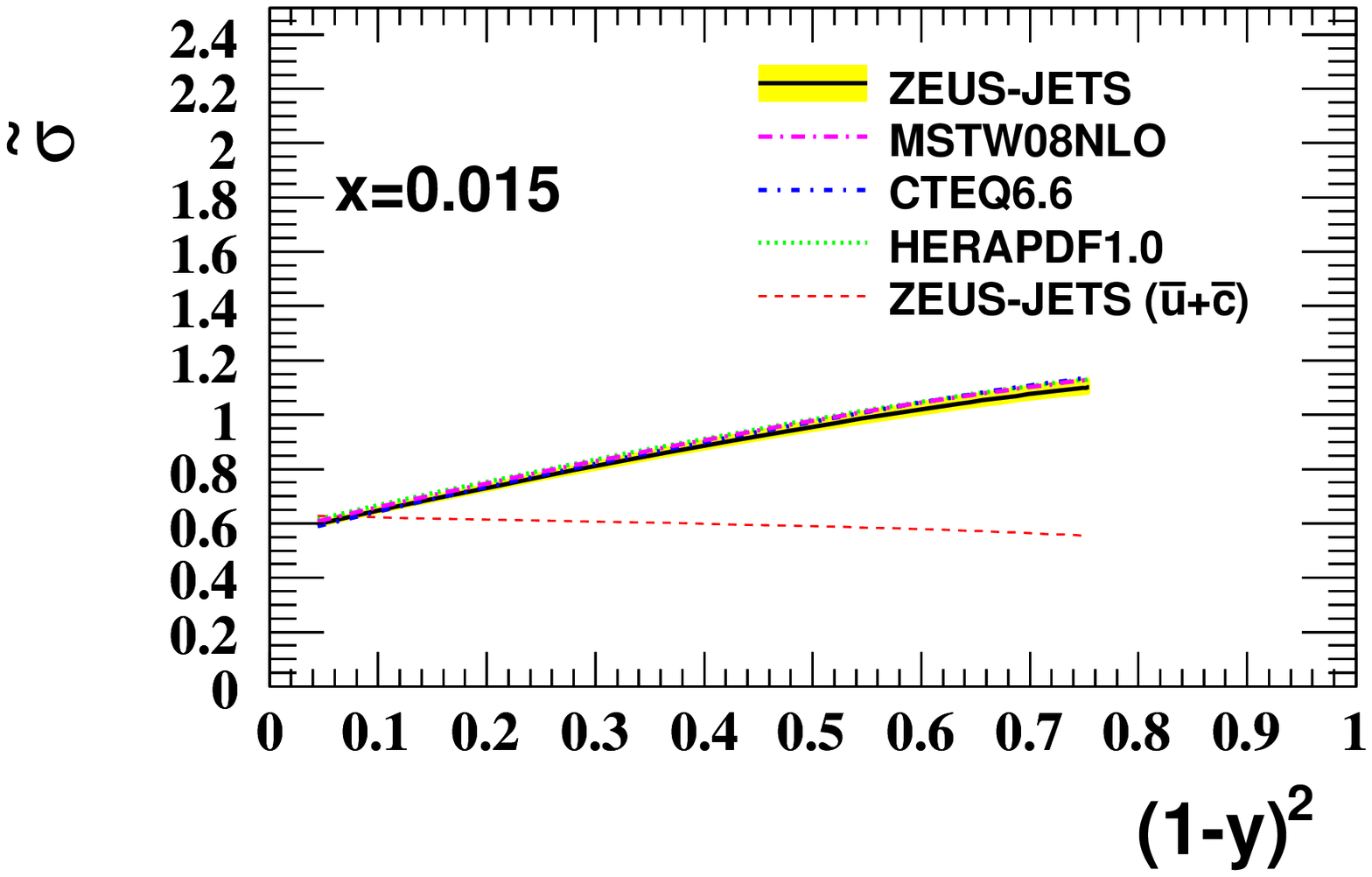}, \includegraphics[width=9.0cm]{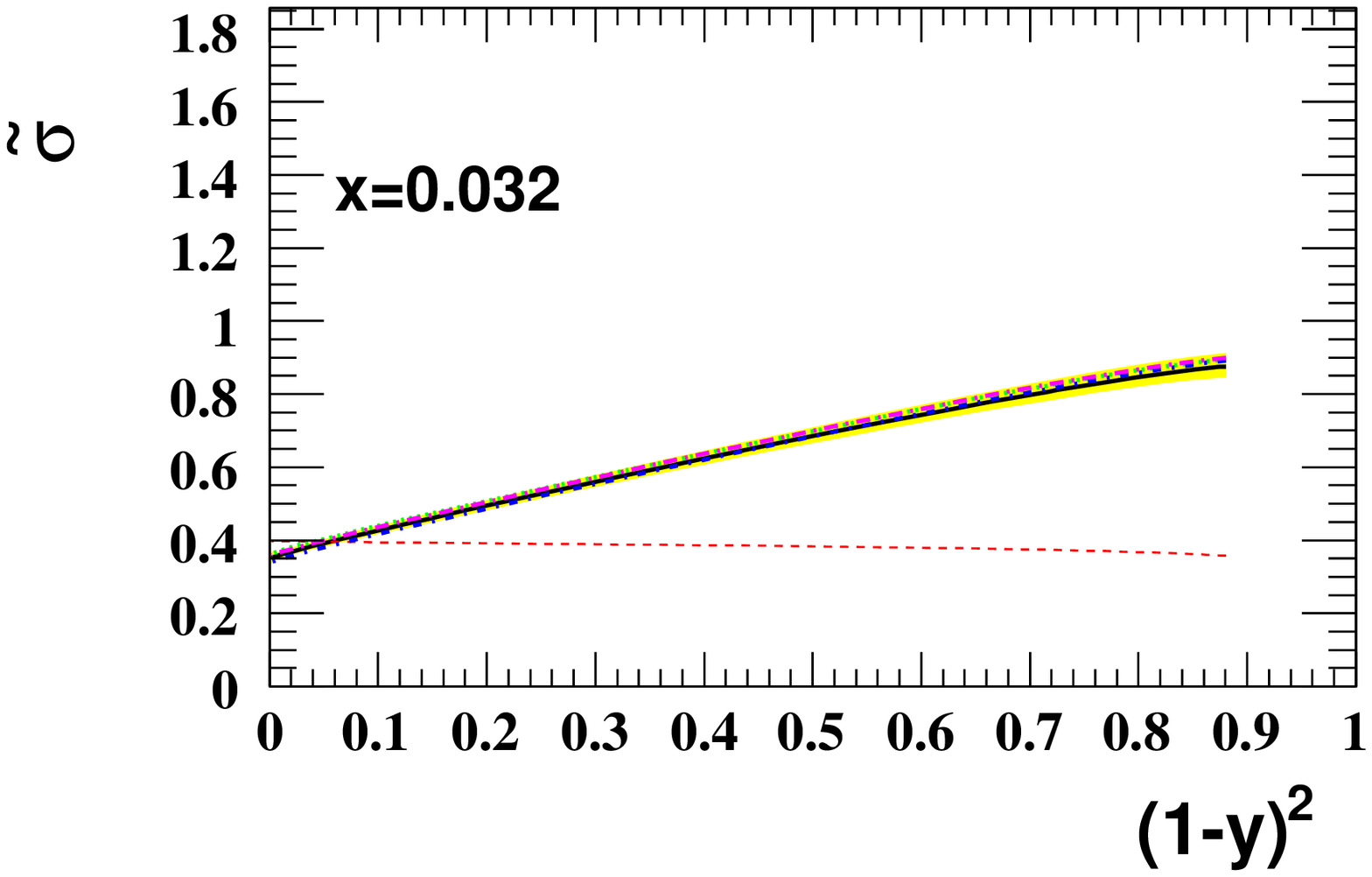}\\
\includegraphics[width=9.0cm]{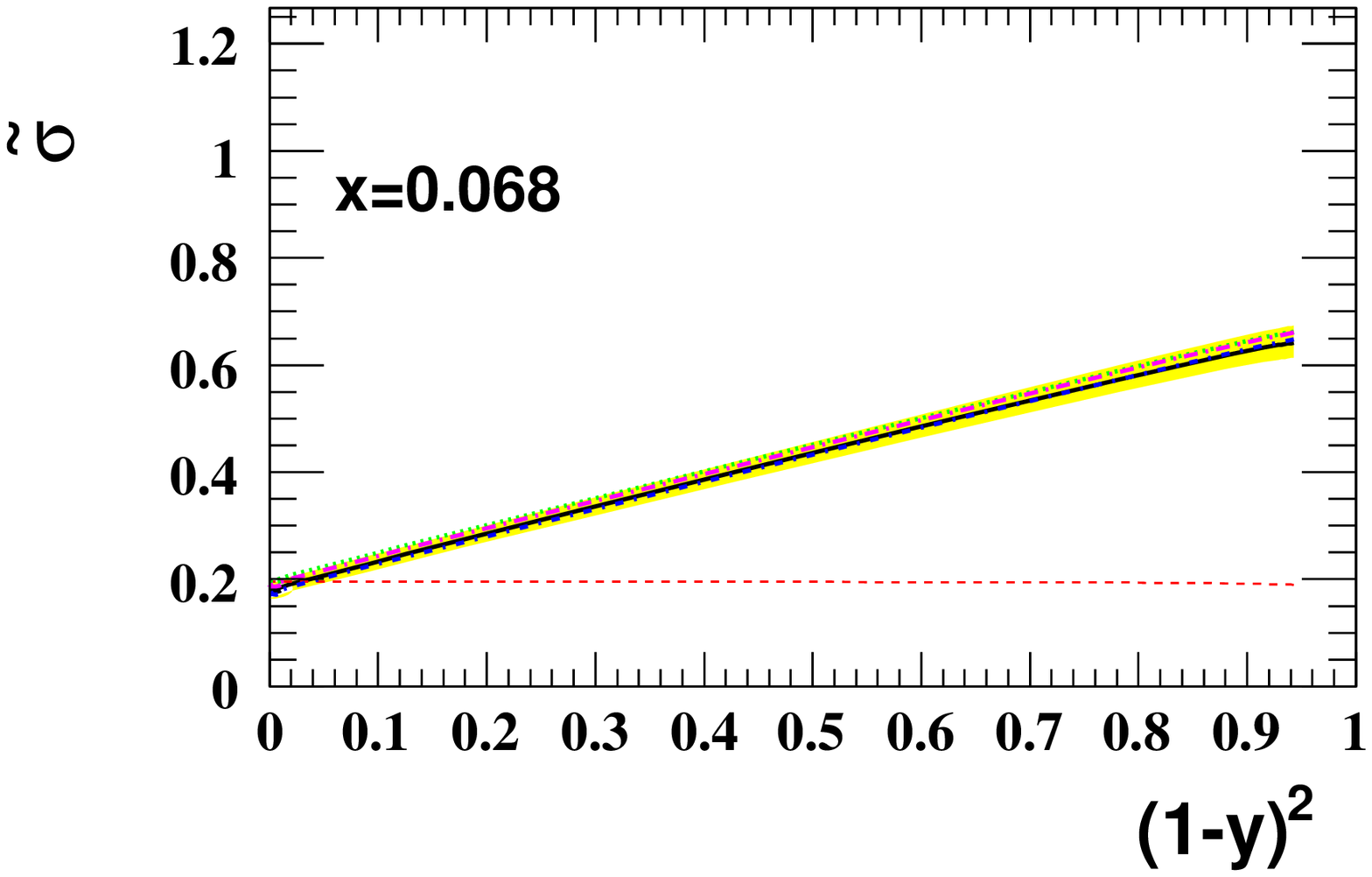}, \includegraphics[width=9.0cm]{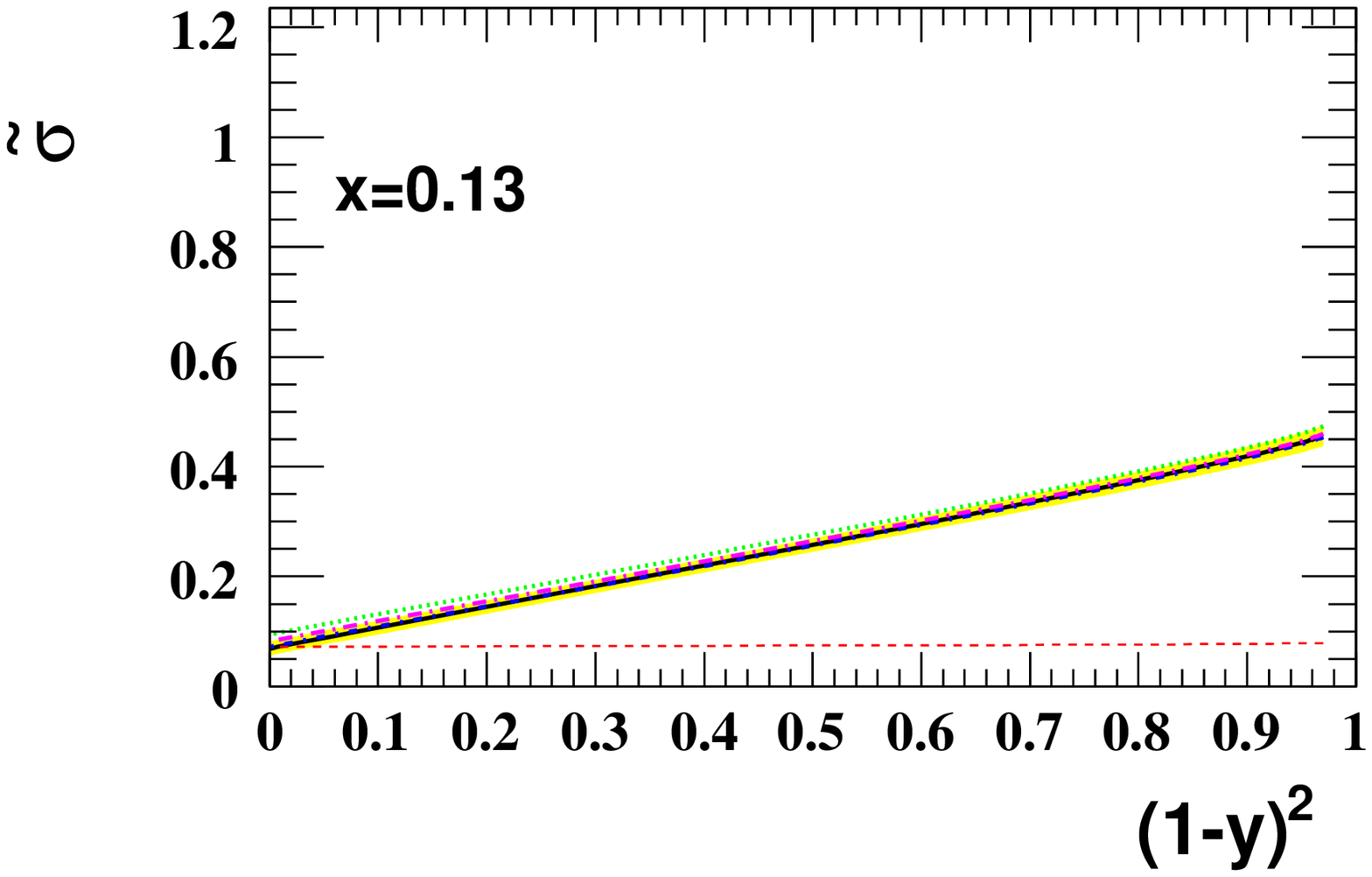}\\
\includegraphics[width=9.0cm]{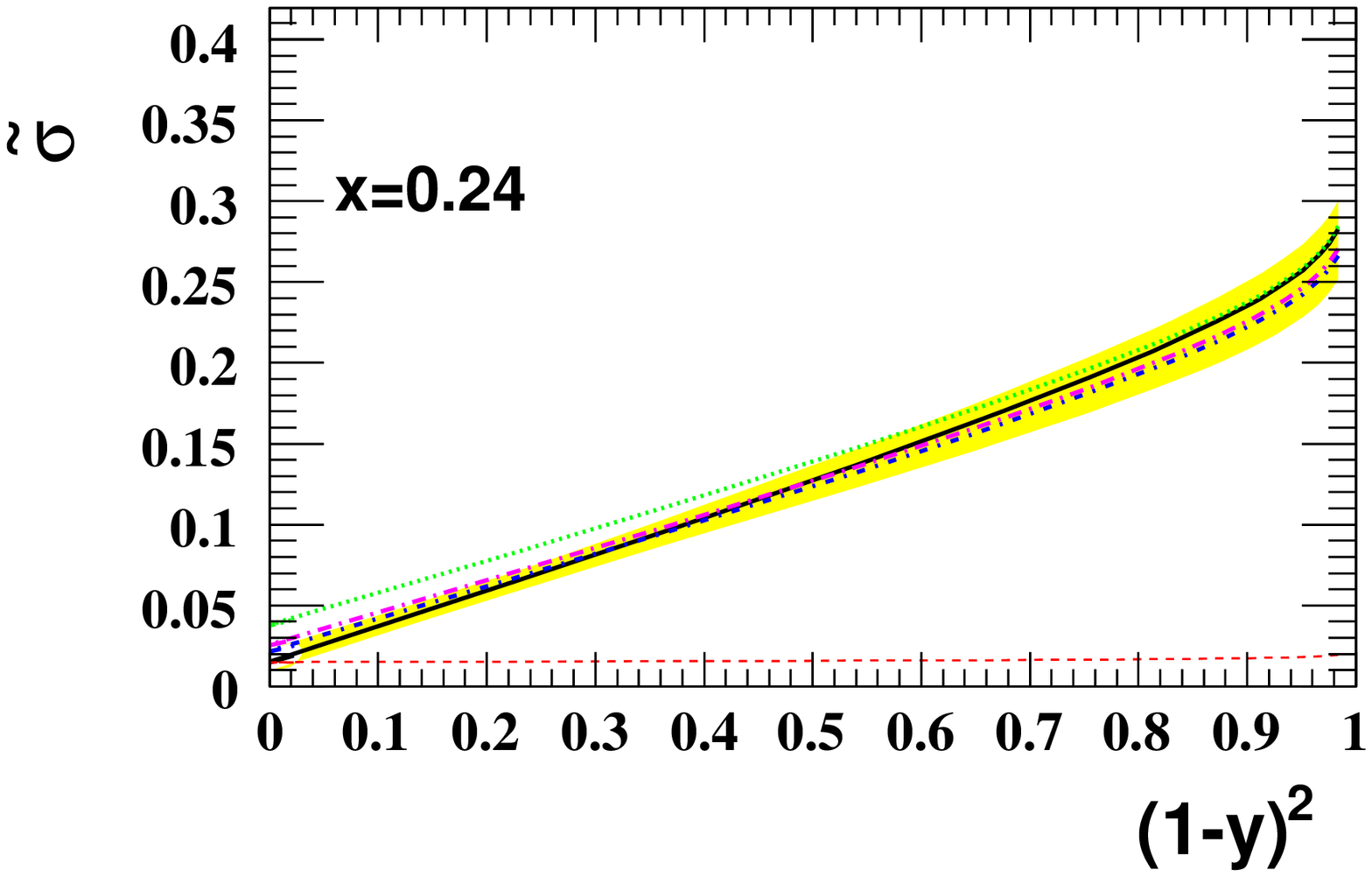}, \includegraphics[width=9.0cm]{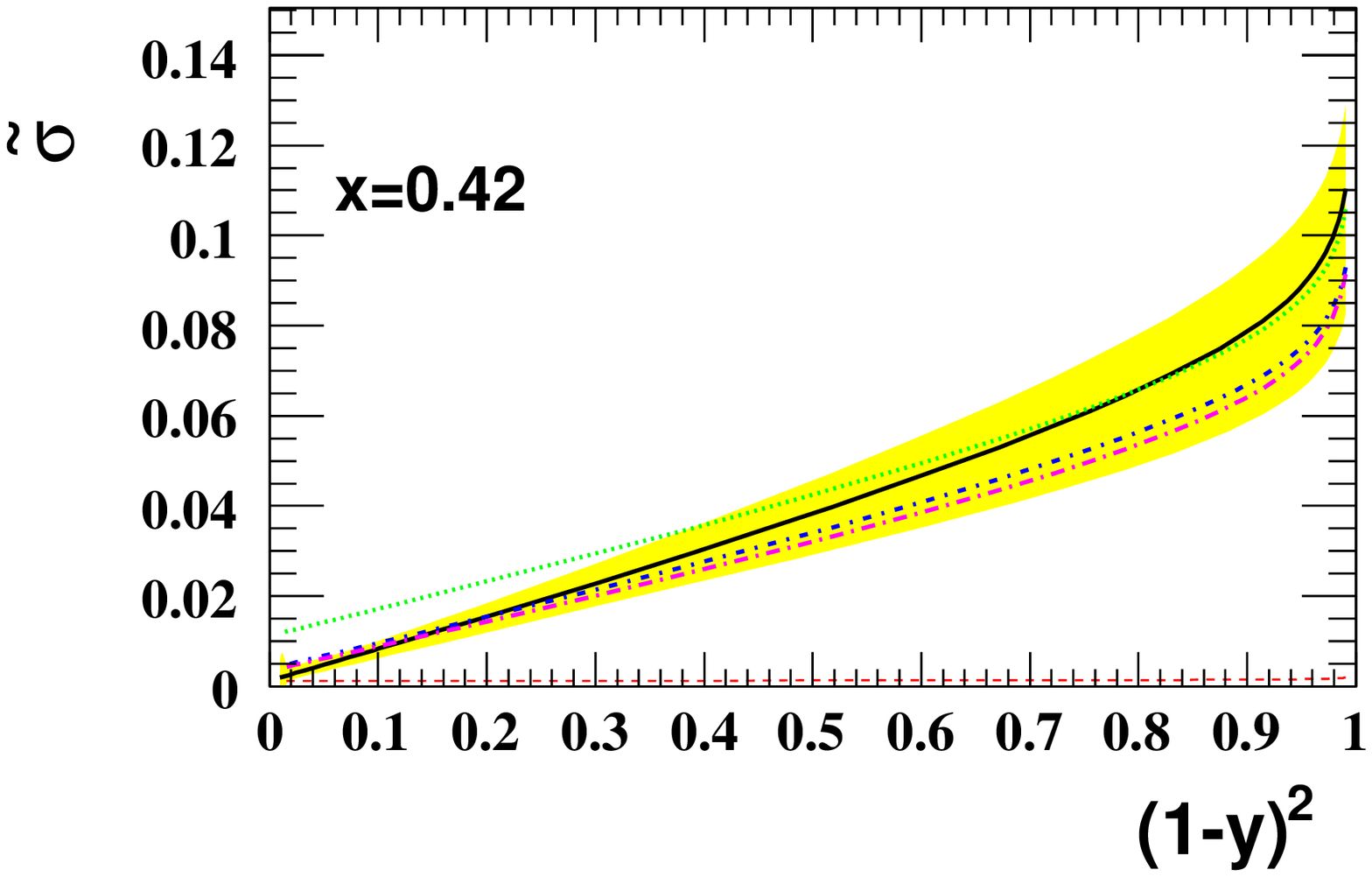}\\
\caption{Predictions of $\tilde{\sigma}$ at NLO in QCD for CC DIS in $e^+p$ collisions with proton energy 920 GeV and positron energy 27.56 GeV. \label{fig:ccpred}}
\end{figure}

Predicted NLO reduced cross sections in CC DIS made using DISPred for $e^+p$ collisons with proton energy 920 GeV and positron energy 27.56 GeV are shown in figure~\ref{fig:ccpred}. Predictions are shown for the PDF sets ZEUS-JETS~\cite{Chekanov:2005nn}, MSTW08~\cite{Martin:2009iq}, CTEQ66~\cite{Tung:2006tb} and HERAPDF1.0~\cite{:2009wt}. In addition the uncertainties for the ZEUS-JETS predictions are shown as a yellow band.

\subsection{Single Differential and Total Cross Sections}

The single and total cross sections at NLO are calculated in precisely the same manner as for the LO calculations except that the NLO expressions for the structure functions are used.

\section{Installation and Usage}
\subsection{Requirements}
The code has been tested on GNU/Linux systems and as such the following packages are required for installation:

\begin{itemize}
\item GNU autoconf automake
\item The GNU scientific library
\item LHAPDF
\item ROOT

\end{itemize}
\subsection{Installation}

Tarballs of the package may be downloaded from 
{\tt \href{http://www.hepforge.org/downloads/dispred}{http://www.hepforge.org/downloads/dispred}}.
After downloadig the tarball, the package may be installed with:\\
{\tt tar -zxvf DISPred-1.0.tgz\\
cd DISPred\\
./configure --prefix=<installation directory>\\
make\\
make install\\
}
This assumes that {\tt root-config} and {\tt lhapdf-config} are already in your path. Do not choose {\tt <installation directory>} to be the same as the location of the expanded tar file. It is now possible to run the example program or to include the DISPred libraries in your own code.

\subsection{Example Program: DISPrediction}

Within  {\tt <installation directory>/bin} there is a program {\tt DISPrediction} which can be used to produce predictions for $ep$ DIS cross sections.
This program takes as input a cards file e.g.:\\
{\tt DISPrediction example.cards}\\

An example set for testing is available in the {\tt example} subdirectory of the tarball. The available options for the cards file are summarised in 
table \ref{tab:cards}. 

\begin{table}
\begin{tabular}{|c|c|c|c|}
\hline 
Type & Name & Default Value & Meaning \\
\hline
  {\tt int}  & {\tt ElectronCharge} & 1& choose $e^+$ or $e^-$ \\
  {\tt int}  & {\tt VegasCalls"} & 50000& Calls to Vegas for integration \\
  {\tt string}& {\tt PDFSetFileName} & "ZEUS2005\_ZJ.LHpdf"& Name of PDF file to use \\
  {\tt int}    & {\tt PDFSubSet} & 0& Subset of PDF to use \\
  {\tt string} & {\tt QCDCalculationLevel} & "LO"& Can be ``LO'' or  ``NLO'' \\
{\tt string} & {\tt DISProcess}  & "NC" & Choose CC or NC \\

  {\tt double} &{\tt ZBosonMass} & 91.1876& $M_Z$ (GeV) \\
  {\tt double} &{\tt WBosonMass} & 80.398&   $M_W$ (GeV) \\
  {\tt double} &{\tt AlphaEM}  & 7.297352570$\times 10^-3$& $\alpha_{\mathrm{EM}}$ \\
  {\tt double} &{\tt GFermi} & 1.16637$10^{-5}$&  $G_F$ \\
  {\tt double} &{\tt TopMass} & 171.2&  $M_t$ \\
  {\tt double} & {\tt BottomMass} & 4.20&  $M_b$ \\

  {\tt double} & {\tt Vub} & 41.2e-3& CKM $V_{ub}$ \\
  {\tt double} &{\tt Vcb}  & 3.93e-3& CKM $V_{cb}$ \\

  {\tt double} & {\tt Sin2ThetaW} & 0.22308&  $\sin^2 \theta_W$  \\
  {\tt double} & {\tt Sin2ThetaC} & 0.05&   $\sin^2 \theta_C$ \\
  {\tt double} & {\tt CouplingVu} & 0.203& $v_u$  SM$=0.5 -4*\sin^2\theta_W$ \\
  {\tt double} & {\tt CouplingVd} & -0.351& $v_d$  SM$=-0.5 +2*\sin^2\theta_W$ \\
  {\tt double} & {\tt CouplingVe} & -0.00538& $v_e$  SM=$-0.5+2*\sin^2\theta_W$ \\

  {\tt double} & {\tt CouplingAu} & 0.5& $a_u$  \\
  {\tt double} & {\tt CouplingAd} & -0.5& $a_d$  \\
  {\tt double} & {\tt CouplingAe} & -0.5& $a_e$ \\

  {\tt string} & {\tt ReducedCrossSection} & "OFF"& can be OFF or ON \\
  {\tt string} & {\tt ReducedCrossSectionBins} & "q2xpoints.dat"& file containing points for $\tilde{\sigma}$  \\

  {\tt double} & {\tt DiffBinPointScale} & 1e-6& Fraction of bin width for $\frac{d\sigma}{dQ^2}$ etc.   \\
  {\tt string} &{\tt DSigmaDQ2} & "OFF"& can be OFF or ON \\
  {\tt string} & {\tt DSigmaDQ2Bins} & "q2bins.dat"& file containing bins for \\
  {\tt string} & {\tt DSigmaDQ2Points} & "AUTO"& file with points for  $\frac{d\sigma}{dQ^2}$ \\
  
  {\tt string} & {\tt DSigmaDX} & "OFF"& can be OFF or ON \\
  {\tt string} & {\tt DSigmaDXBins} & "xbins.dat"& file containing bins for $\frac{d\sigma}{dx}$  \\
  {\tt string} & {\tt DSigmaDXPoints}& "AUTO"& file with points for  \\

  {\tt string} & {\tt DSigmaDY} & "OFF"& can be OFF or ON  \\
  {\tt string} & {\tt DSigmaDYBins} & "ybins.dat"& file containing points for $\frac{d\sigma}{dy}$  \\
  {\tt string} & {\tt DSigmaDYPoints} & "AUTO"& file with points for  \\

  {\tt double} & {\tt Q2Min} & 0.0&     minimum $Q^2$ \\
  {\tt double} & {\tt Q2Max} & 100000.0& maximum $Q^2$  \\

  {\tt double} & {\tt XMin} & 0.0& minimum $x$ \\
  {\tt double} & {\tt XMax} & 1.0& maximum $x$ \\

  {\tt double} & {\tt YMin} & 0.0& minimum $y$ \\
  {\tt double} & {\tt YMax} & 1.0& maximum $y$  \\

  {\tt double} & {\tt ELepton} & 27.5&  Electron beam energy \\
  {\tt double} & {\tt EProton} & 920.0& Proton beam energy \\

  {\tt string} & {\tt ROOTOutputFile} & "DISPredOut.root" & Output file for ROOT\\
\hline
\end{tabular}

\caption{Available control cards for {\tt DISPrediction}.\label{tab:cards}}
\end{table}

\subsection{DISPred Library and Classes}

The DISPred packagedprovides a library as well as the {\tt DISPrediction} executable. This library makes it easy to construct programmes that calculate
DIS cross sections. An example of a simple programme is in fact {\tt DISPrediction} itself, which is very short:

{\tt\ \\
\#include <iostream>\\
\#include "DISPredictor.h"\\
\ \\
using namespace DISPred;\\
\ \\
int main (int argc, char **argv)\{\\
 \ std::cout << "DISPrediction v1.0 -  31 Mar 2010" << std::endl;\\
 // Create instance of DISPredictor\\
 \ DISPred::DISPredictor *DISPred= DISPred::DISPredictor::Instance();
 // initialise from control cards provided via command line\\
 \ DISPred->Initialise(argc,argv);
 \ DISPred->CalculateCrossSections();\\
 \ DISPred->PrintResults();\\
 \ std::cout << "DISPrediction v1.0 - Run finished Succesfully" << std::endl;\\
 \ DISPred->WriteOutput();\\
 \ return 0;\\
\}
}

All classes are part of the name space {\tt DISPred}.

\subsubsection{The {\tt ControlCards} Class}

The control cards class is used to handle configuration options that can be 
read in from a text file. It is implemented as a singleton class. Available
methods for the class are detailed below.
\begin{description}
\item {\tt ControlCards* Instance()}: Returns a pointer to
the instance of control cards.
\item {\tt void AddCardDouble(const std::string key, const double defval)}: Defines a card with name {\tt key} and with a default double precision value {\tt defval}.
\item {\tt void AddCardInt(const std::string key, const int defval)}: Defines a card with name {\tt key} and with a default value {\tt defval} which is an integer.
\item {\tt void AddCardString(const std::string key, std::string defval) }: Defines a card with name {\tt key} and with a default value {\tt defval} which is a string.
\item {\tt void AddCardVector(const std::string key, const std::vector<double> defval)}: Defines a card with name {\tt key} and with a default value {\tt defval} which is a vector of double precision values.
\item {\tt int readKeys(const char* fileName)}: Reads in card values from the file with name {\tt fileName}.
\item {\tt double fetchValueDouble(const std::string\& key)}: fetch the value of card {\tt key}.
\item {\tt int fetchValueInt(const std::string\& key)}: fetch the values of card {\tt key}.
\item {\tt std::string fetchValueString(const std::string\& key)}: fetch the value of card {\tt key}.
\item {\tt std::vector<double> fetchValueVector(const std::string\& key)}: fetch the values of card {\tt key}.
\item {\tt void printCards():} Print  current card values to stdout.
\end{description}

\subsubsection{The {\tt DISPredictor} class}

The {\tt DISPredictor} class is a singleton class that is the workhorse of DISPred. It has many public methods.

\begin{description}
\item {\tt static DISPredictor* Instance()}: returns the instance of {\tt DISPredictor}.
\item {\tt void Initialise(int my\_argc, char **my\_argv)}: Initialise DISPredictor based on a cards file name which can come directly from {\tt stdin}.
\item {\tt void CalculateCrossSections()}: Calculate cross sections as configured in the cards.   
\item {\tt void InitPDF(int subset)}: Initialise the chosen PDF set.
\item {\tt void PrintResults()}: Print results to {\tt stdout}.
\item {\tt void WriteOutput()}: Write the output rootfile.
\item {\tt double CalculateReducedCrossSection(double x, double q2)}: Calculate a NC DIS reduced cross section.
\item {\tt double CalculateCCReducedCrossSection(double x, double q2)}: Calculate a CC DIS reduced cross section.
\item {\tt double CalculatePropagator(double q2, double x)}: Calculate the NC propagator.
\item {\tt double CalculateCCPropagator(double q2, double x)}: Calculate the CC propagator.
\item {\tt double CalculateDSigmaDQ2(double q2min,double q2max)}: Calculate $\frac{d\sigma}{dQ^2}$.
\item {\tt double CalculateDSigmaDX(double xmin,double xmax)}: Calculate $\frac{d\sigma}{dx}$.
\item {\tt double CalculateQ2DSigmaDQ2(double q2min,double q2max)}: Calculate $Q^2\frac{d\sigma}{dQ^2}$.
\item {\tt double CalculateXDSigmaDX(double xmin,double xmax)}:Calculate $x\frac{d\sigma}{dx^2}$.
\item {\tt double CalculateDSigmaDY(double ymin,double ymax)}: Calculate $\frac{d\sigma}{dy}$.
\item {\tt double CalculateYDSigmaDY(double ymin,double ymax)}: Calculate $y\frac{d\sigma}{dy}$.
\item {\tt double S()}: Return the centre-of-mass energy squared.

\end{description}

\subsubsection{The {\tt RedSigmaPoint} class}
The {\tt RedSigmaPoint} class is a simple class for storing information about
double-differential cross sections points. For each point the 
$Q^2$({\tt \_q2}), $x$({\tt \_x}), $\tilde{\sigma}$({\tt \_redsigma}) and $\frac{d^2 \sigma}{dQ^2 dx}$ ({\tt \_d2sdq2dx}). 
\begin{description}
\item {\tt RedSigmaPoint(double q2, double x)}: constructor that creates a point with {\tt \_q2}={\tt q2} and {\tt \_x}={\tt x} and other values 0. 
\item {\tt RedSigmaPoint(double q2, double x,double redsigma)}: constructor that creates a point with {\tt \_q2}={\tt q2} and {\tt \_x}={\tt x}, {\tt \_redsigma}={\tt redsigma} and {\tt \_d2sdq2dx}=0.
\item {\tt RedSigmaPoint(double q2, double x,double redsigma, double d2sdq2dx )}: constructor that creates a point with {\tt \_q2}={\tt q2} and {\tt \_x}={\tt x}, {\tt \_redsigma}={\tt redsigma} and {\tt \_d2sdq2dx}={\tt d2sdq2dx}.
\item {\tt RedSigmaPoint()}: Constructor with all vlaues set to 0;
\item {\tt double Q2()}: returns {\tt \_q2}.
\item {\tt double X()}: returns {\tt \_x}.
\item {\tt double RedSigma()}: returns {\tt \_redsigma}.
\item {\tt double D2sDQ2Dx()}: returns {\tt \_d2sdq2dx}.
\item {\tt void SetRedSigma(double reduced)}: Set {\tt \_redsigma}.
\item {\tt void SetD2sDQ2Dx(double reduced)}: Set {\tt \_d2sdq2dx}.
\item {\tt void Print()}: Print out information.
\item {\tt void PrintShort()}: Briefly print out information.
\end{description}

\subsubsection{The {\tt RedSigmaGrid} class}

The {\tt RedSigmaGrid} class inherits from a {\tt std::vector<RedSigmaPoint>}. With the following extra methods:
\begin{description}
\item {\tt void Print()}: Print out information.
\item {\tt void PrintShort()}: Briefly print out information.
\end{description}

\subsubsection{The {\tt DiffSigmaPoint} class}
The {\tt DiffSigmaPoint} class is a simple class for storing information about
single-differential cross sections at a point. A point in the variable of choice called {\tt \_var} and the differential cross section {\tt \_diffsigma} are stored. The following public methods are available.
\begin{description}
\item {\tt DiffSigmaPoint()}: Default constructor, sets {\tt \_var} to 1.5 and {\tt \_diffsigma} to 0.
\item {\tt DiffSigmaPoint(double var)}: Constructor that creates a point at {\tt \_var}= {\tt var} with {\tt \_diffsigma}=0.
\item {\tt DiffSigmaPoint(double var, double diffsigma)}: Constructor that creates a point at {\tt \_var}= {\tt var} with {\tt \_diffsigma}={\tt diffsigma}.
\item {\tt void Print()}: Print out information.
\item {\tt void PrintShort()}: Briefly print out information.
\item {\tt double Var()}: returns {\tt \_var}.
\item {\tt double DiffSigma()}: returns {\tt \_diffsigma}.
\item {\tt void SetDiffSigma(double diffsigma)}: sets {\tt \_diffsigma} to {\tt diffsigma}.
\end{description}

\section{Root Output}
When DISPred produces an output root file, then a {\tt TTree} and several histograms and graphs are produced.
\subsection{Root {\tt TTree}}
A {\tt TTree } called {\tt ReducedCrossSections} is produced. The variables in this tree are listed in table \ref{tab:treevars}.

\begin{table}
\begin{center}
\begin{tabular}{|c|c|c|}
\hline
Type & Variable Name & Description \\
\hline
{\tt int} &  {\tt point}   & An integer giving the ID of the point        \\
{\tt double} &  {\tt Q2}  & The $Q^2$ of the point         \\
{\tt double} &  {\tt x}  & The $x$ of the point         \\
{\tt double} &  {\tt ddiffsigma}  & The double-differential cross section  \\
{\tt double} &  {\tt redsigma}  & The reduced cross section  \\
{\tt double} &  {\tt d}  & The $d$ PDF at this point  \\
{\tt double} &  {\tt dbar}  & The $\bar{d}$ PDF at this point  \\
{\tt double} &  {\tt u}  & The $u$ PDF at this point  \\
{\tt double} &  {\tt ubar}  & The $\bar{u}$ PDF at this point  \\
{\tt double} &  {\tt s}  & The $s$ PDF at this point  \\
{\tt double} &  {\tt sbar}  & The $\bar{s}$ PDF at this point  \\
{\tt double} &  {\tt c}  & The $c$ PDF at this point  \\
{\tt double} &  {\tt cbar}  & The $\bar{c}$ PDF at this point  \\
{\tt double} &  {\tt b}  & The $b$ PDF at this point  \\
{\tt double} &  {\tt bbar}  & The $\bar{b}$ PDF at this point  \\
\hline
\end{tabular}
\end{center}
\caption{Tree variables in the root output file.\label{tab:treevars}}
\end{table}

\subsection{Root Histograms}

Six {\tt TH1D} objects are produced:
\begin{description}
\item {\tt DSigmaDQ2}: Binwise ${\frac{d\sigma}{dQ^2}}$;
\item {\tt DSigmaDX}: Binwise ${\frac{d\sigma}{dx}}$;
\item {\tt DSigmaDY}: Binwise ${\frac{d\sigma}{dy}}$.
\item {\tt Q2DSigmaDQ2}: Binwise $Q^2{\frac{d\sigma}{dQ^2}}$;
\item {\tt XDSigmaDX}: Binwise $x{\frac{d\sigma}{dx}}$;
\item {\tt YDSigmaDY}: Binwise $y{\frac{d\sigma}{dy}}$.
\end{description}

\subsection{Root Graphs}

Three {\tt TGraphAsymErrors} objecs are produced:
\begin{description}
\item {\tt GraphDSigmaDQ2}: Pointwise ${\frac{d\sigma}{dQ^2}}$;
\item {\tt GraphDSigmaDX}: Pointwise ${\frac{d\sigma}{dx}}$;
\item {\tt GraphDSigmaDY}: Pointwise ${\frac{d\sigma}{dy}}$.
\end{description}

\section{Summary}

This manual for the DISPred package v1.0 has outlined the features currently
implemented together with a simple example programme that will make 
predictions for DIS cross sections in $ep$ scattering. The code and most up-to-date information are hosted by hepforge at:  {\tt \href{http://projects.hepforge.org/dispred/}{http://projects.hepforge.org/dispred/}}.

\section*{Acknowledgements}
The author wishes to thank C.~Gwenlan for help with {\sc QCDNUM} and cross
 checks of the results from DISPred, M.~Sutton for testing the code, A.~Tapper for providing code that is used for the implementation of control cards and  R.~Ciesielski,  A.~Cooper-Sarkar, K.~Oliver, E.~Tassi and M.~Turcato for feedback on the results.


\providecommand{\href}[2]{#2}\begingroup\raggedright\endgroup

\end{document}